\documentclass{elsart}

\usepackage{graphics}
\usepackage{epsfig}
\usepackage{subfigure}
\usepackage{amsmath,amsfonts,amssymb,graphicx,times}

\begin{document}

\begin{frontmatter}

\title{Effects of high-order correlations on personalized recommendations for bipartite networks}

\author[1,2,3]{Jian-Guo Liu}\ead{liujg004@ustc.edu.cn},
\author[1,2,3]{Tao Zhou}\ead{zhutou@ustc.edu},
\author[1]{Hong-An Che},
\author[1,3]{Bing-Hong Wang},
\author[1,2,3]{and Yi-Cheng Zhang}

\address[1]{Research Center of Complex Systems Science, University of
Shanghai for Science and Technology, Shanghai 200093, PR China}
\address[2]{Department of Physics, University of Fribourg, Chemin du
Mus\'{e}e 3, CH-1700 Fribourg, Switzerland}
\address[3]{Department of Modern Physics, University of Science and Technology of China, Hefei 230026, PR
China}

\begin{abstract}
In this paper, we introduce a modified collaborative filtering (MCF)
algorithm, which has remarkably higher accuracy than the standard
collaborative filtering. In the MCF, instead of the cosine
similarity index, the user-user correlations are obtained by a
diffusion process. Furthermore, by considering the second-order
correlations, we design an effective algorithm that depresses the
influence of mainstream preferences. Simulation results show that
the algorithmic accuracy, measured by the average ranking score, is
further improved by 20.45\% and 33.25\% in the optimal cases of
MovieLens and Netflix data. More importantly, the optimal value
$\lambda_{\rm opt}$ depends approximately monotonously on the
sparsity of the training set. Given a real system, we could estimate
the optimal parameter according to the data sparsity, which makes
this algorithm easy to be applied. In addition, two significant
criteria of algorithmic performance, diversity and popularity, are
also taken into account. Numerical results show that as the sparsity
increases the algorithm considered the second-order correlation can
outperform the MCF simultaneously in all three criteria.
\end{abstract}


\end{frontmatter}

\section{Introduction}
With the expansion of the Internet services, people are becoming
increasingly dependent on the Internet with an information overload.
Consequently, how to efficiently help people find information that
they truly need is a challenging task nowadays \cite{Resnkck1997}.
Being an effective tool to address this problem, the recommender
system has caught increasing attention and become an essential issue
in Internet applications such as e-commerce system and digital
library system \cite{Ecommerce2001}. Motivated by the practical
significance to the e-commerce and society, the design of an
efficient recommendation algorithm becomes a joint focus from
engineering science to mathematical and physical community. Various
kinds of algorithms have been proposed, such as correlation-based
methods \cite{Herlocker2004,Konstan1997}, content-based methods
\cite{Balab97,Pazzani99,Gao2009,Luo2009}, spectral analysis
\cite{Billsus1998,Sarwar2000a}, iteratively self-consistent
refinement \cite{Jie2008}, principle component analysis
\cite{Goldberg2001}, network-based methods
\cite{Zhang2007a,Zhang2007b,Zhou2007,Zhou2007b},  and so on. For a
review of current progress, see Ref. \cite{Adomavicius2005,Liu2009b}
and the references therein.

One of the most successful recommendation algorithms, called
\emph{collaborative filtering} (CF), has been developed and
extensively investigated over the past decade
\cite{Herlocker2004,Konstan1997,Duo2009}. When predicting the
potential interests of a given user, such approach firstly
identifies a set of similar users from the past records and then
makes a prediction based on the weighted combination of those
similar users' opinions. Despite its wide applications,
collaborative filtering suffers from several major limitations
including system scalability and accuracy \cite{Sarwar2000}.
Recently, some physical dynamics, including mass diffusion (MD)
\cite{Zhang2007b,Zhou2007,Liu2009} and heat conduction (HC)
\cite{Zhang2007a}, have found their applications in personalized
recommendations. Based on MD and HC, several effective network-based
recommendation algorithms have been proposed
\cite{Zhang2007a,Zhang2007b,Zhou2007,Zhou2007b}. These algorithms
have been demonstrated to be of both high accuracy and low
computational complexity. However, the algorithmic accuracy and
computational complexity may be very sensitive to the statistics of
data sets. For example, the algorithm presented in Ref.
\cite{Zhou2007} runs much faster than the standard CF if the number
of users is much larger than that of objects, while when the number
of objects is huge, the advantage of this algorithm vanishes because
its complexity is mainly determined by the number of objects (see
Ref. \cite{Zhou2007} for details). Since the CF algorithm has been
extensively applied in the real e-commerce systems
\cite{Konstan1997,CF2}, it's meaningful to find some ways to
increase the algorithmic accuracy of CF. We therefore present a
modified collaborative filtering (MCF) method, in which the user
correlation is defined based on the diffusion process. Recently, Liu
{\it et al.} \cite{Liu2009d} studied the user and object degree
correlation effect to CF, they found that the algorithm accuracy
could be remarkably improved by adjusting the user and object degree
correlation. In this paper, we argue that the high-order
correlations should be taken into account to depress the influence
of mainstream preferences and the accuracy could be improved in this
way. The correlation between two users is, in principle, an
integration of many underlying similar tastes. For two arbitrary
users, the very specific yet common tastes shall contribute more to
the similarity measure than those mainstream tastes. Figure 1 shows
an illustration of how to find the specific tastes by eliminating
the mainstream preference. To the users $A$ and $C$, the commonly
selected objects 1 and 2 could reflect their tastes, where 1 denotes
the mainstream preference shared by all $A$, $B$ and $C$, and 2 is
the specific taste of $A$ and $C$. Both 1 and 2 contribute to the
correlation between $A$ and $C$. Since 1 is the mainstream
preference, it also contributes to the correlations between $A$ and
$B$, as well as $B$ and $C$. Tracking the path $A\rightarrow B
\rightarrow C$, the mainstream preference 1 could be identified by
considering the second-order correlation between $A$ and $C$.
Statistically speaking, two users sharing many mainstream
preferences should have high second-order correlation, therefore we
can depress the influence of mainstream preferences by taking into
account the second-order correlation. The numerical results show
that the algorithm involving high-order correlations is much more
accurate and provides more diverse recommendations.

\begin{figure}
\center\scalebox{0.5}[0.5]{\includegraphics{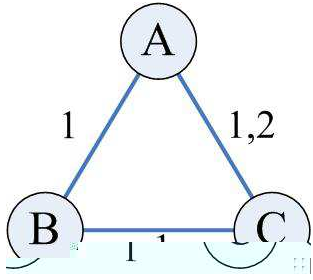}}
\caption{Illustration of the user correlation network. The users
$A$, $B$ and $C$ are correlated because they have collected some
common objects, where object $1$ has been collected by all of the
three users, while object 2 is only collected by user $A$ and $C$.
}\label{Fig1}
\end{figure}

\section{Problem description and performance metrics}

Denote the object set as $O = \{o_1,o_2, \cdots, o_m\}$ and the user
set as $U$ = $\{u_1, u_2,$ $\cdots,$  $u_n\}$, a recommender system
can be fully described by an adjacent matrix $A=\{a_{ij}\}\in
R^{m,n}$, where $a_{ij}=1$ if $o_i$ is collected by $u_j$, and
$a_{ij}=0$ otherwise. For a given user, a recommendation algorithm
generates an ordered list of all the objects he/she has not
collected before.

To test the recommendation algorithmic accuracy, we divide the data
set into two parts: one is the training set used as known
information for prediction, and the other one is the probe set,
whose information is not allowed to be used. Many metrics have been
proposed to judge the algorithmic accuracy, including
\emph{precision} \cite{Adomavicius2005}, \emph{recall}
\cite{Adomavicius2005}, \emph{F-measure} \cite{Herlocker2004},
\emph{average ranking score} \cite{Zhou2007}, and so on. Since the
average ranking score does not depend on the length of
recommendation list, we adopt it in this paper. Indeed, a
recommendation algorithm should provide each user with an ordered
list of all his/her uncollected objects. For an arbitrary user
$u_i$, if the entry $u_i$-$o_j$ is in the probe set (according to
the training set, $o_j$ is an uncollected object for $u_i$), we
measure the position of $o_j$ in the ordered list. For example, if
there are $L_i=100$ uncollected objects for $u_i$, and $o_j$ is the
10th from the top, we say the position of $o_j$ is $10/100$, denoted
by $r_{ij}=0.1$. Since the probe entries are actually collected by
users, a good algorithm is expected to give high recommendations,
leading to small $r_{ij}$. Therefore, the mean value of the position
$r_{ij}$, $\langle r\rangle$ (called \emph{average ranking score}
\cite{Zhou2007}), averaged over all the entries in the probe, can be
used to evaluate the algorithmic accuracy: the smaller the ranking
score, the higher the algorithmic accuracy, and vice verse. For a
null model with randomly generated recommendations, $\langle
r\rangle=0.5$.

Besides accuracy, the average degree of all recommended objects,
$\langle k\rangle$, and the mean value of Hamming distance, $S$, are
taken into account to measure the algorithmic popularity and
diversity \cite{Zhou2007b}. The smaller average degree,
corresponding to the less popular objects, are preferred since those
lower-degree objects are hard to be found by users themselves. In
addition, the personal recommendation algorithm should present
different recommendations to different users according to their
tastes and habits. The diversity can be quantified by the average
Hamming distance, $S=\langle H_{ij}\rangle$, where
$H_{ij}=1-Q_{ij}/L$, $L$ is the length of recommendation list, and
$Q_{ij}$ is the overlapped number of objects in $u_i$'s and $u_j$'s
recommendation lists. The higher $S$ indicates a more diverse and
thus more personalized recommendations.

\section{Modified collaborative filtering algorithm based on diffusion process}
In the standard CF, the correlation between $u_i$ and $u_j$ can be
evaluated directly by the well-known cosine similarity index
\begin{equation}\label{equationa}
s_{ij}^c=\frac{\sum_{l=1}^ma_{li}a_{lj}}{\sqrt{k(u_i)k(u_j)}},
\end{equation}
where $k(u_i)=\sum_{l=1}^ma_{li}$ is the degree of user $u_i$.
Inspired by the diffusion process presented by Zhou {\it et al.}
\cite{Zhou2007}, the user correlation network can be obtained by
projecting the user-object bipartite network. How to determine the
edge weight is the key issue in this process. We assume a certain
amount of resource (e.g., recommendation power) is associated with
each user, and the weight $s_{ij}$ represents the proportion of the
resource $u_j$ would like to distribute to $u_i$. This process could
be implemented by applying the network-based resource-allocation
process \cite{Ou2007} on a user-object bipartite network where each
user distributes his/her initial resource equally to all the objects
he/she has collected, and then each object sends back what it has
received to all the users who collected it, the weight $s_{ij}$ (the
fraction of initial resource $u_j$ eventually gives to $u_i$) can be
expressed as:
\begin{equation}\label{equation2}
s_{ij}=\frac{1}{k(u_j)}\sum^m_{l=1}\frac{a_{li}a_{lj}}{k(o_l)},
\end{equation}
where $k(o_l)=\sum^n_{i=1}a_{li}$ denotes the degree of object
$o_l$. For the user-object pair $(u_i,o_j)$, if $u_i$ has not yet
collected $o_j$ (i.e., $a_{ji}=0$), the predicted score, $v_{ij}$,
is given as
\begin{equation}\label{equation1}
v_{ij}=\frac{\sum_{l=1}^ns_{li}a_{jl}}{\sum_{l=1}^ns_{li}}.
\end{equation}
Based on the definitions of $s_{ij}$ and $v_{ij}$, given a target
user $u_i$, the MCF algorithm is given as following
\begin{description}
\item[(i)] Calculating the user correlation matrix $\{s_{ij}\}$ based
on the diffusion process, as shown in Eq. (2);
\item[(ii)] For each user $u_i$, based on Eq. (3), calculating the
predicted scores for his/her uncollected objects;
\item[(iii)] Sorting
the uncollected objects in descending order of the predicted scores,
and those objects in the top will be recommended.
\end{description}
The standard CF and the MCF have similar process, and their only
difference is that they adopt different measures of user-user
correlation (i.e., $s_{ij}^c$ for the standard CF and $s_{ij}$ for
MCF).


\begin{figure}
\center\scalebox{0.4}[0.4]{\includegraphics{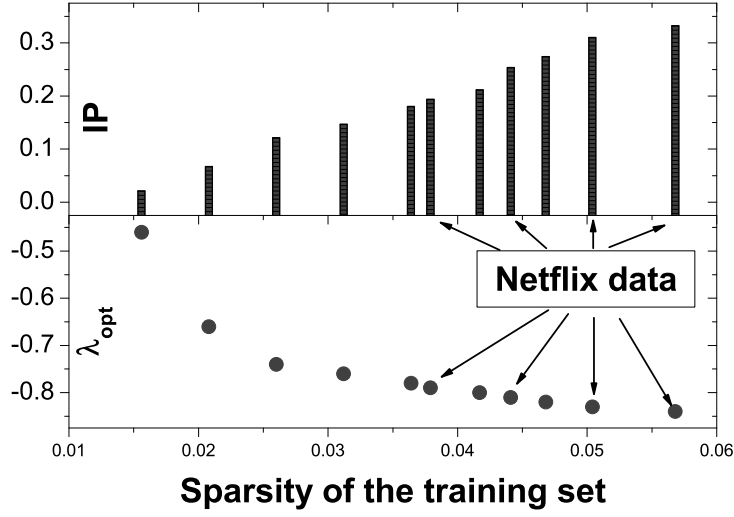}}
\caption{The optimal $\lambda_{\rm opt}$ and the improvement (IP)
vs. the sparsity of the training sets. All the data points are
averaged over ten independent runs with different data-set
divisions. The results corresponding to Netflix data are
marked.}\label{Fig24}
\end{figure}

\section{Numerical results of MCF}

We use two benchmark data sets, one is \emph{MovieLens}\footnote{
http://www.grouplens.org}, which consists of 1682 movies (objects)
and 943 users. The other one is
\emph{Netflix}\footnote{http://www.netflixprize.com}, which consists
of 3000 movies and 3000 users (we use a random sample of the whole
Netflix dataset). The users vote movies by discrete ratings from one
to five. Here we applied a coarse-graining method
\cite{Zhou2007,Zhou2007b}: A movie is set to be collected by a user
only if the giving rating is larger than 2. In this way, the
\emph{MovieLens} data has 85250 edges, and the \emph{Netflix} data
has 567456 edges. The data sets are randomly divided into two parts:
the training set contains $p$ percent of the data, and the remaining
$1-p$ part constitutes the probe.

Implementing the standard CF and MCF when $p=0.9$, the average
ranking scores on \emph{MovieLens} and \emph{Netflix} data are
improved from from 0.1168 to 0.1038 and from 0.2323 to 0.2151,
respectively. Clearly, using the simply diffusion-based simlarity,
subject to the algorithmic accuracy, the MCF outperforms the
standard CF. The corresponding average object degree and diversity
are also improved (see Fig.\ref{Fig22} and Fig.\ref{Fig23} below).

\begin{figure}
\center\scalebox{0.4}[0.4]{\includegraphics{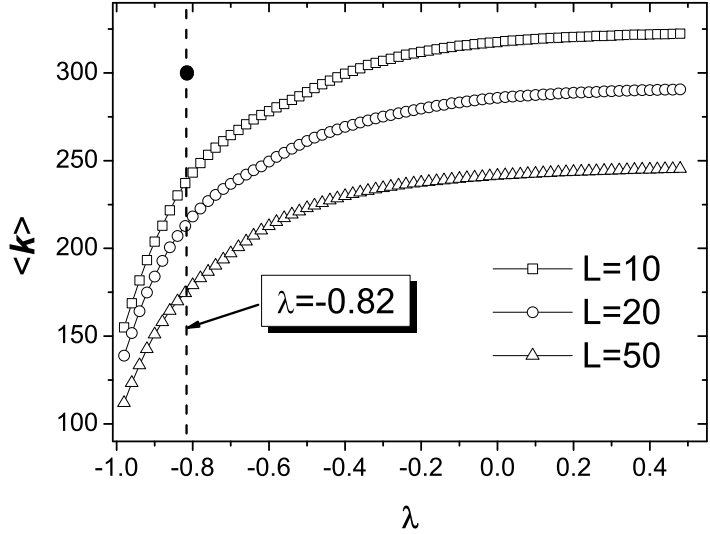}}
\caption{Average degree of recommended objects, $\langle k\rangle$,
vs. $\lambda$ when $p=0.9$. Squares, circles and triangles represent
lengths $L=10, 20$ and $50$, respectively. The black point
($\bullet$) corresponds to the average degree obtained by the
standard CF with $L=20$. All the data points are averaged over ten
independent runs with different data-set divisions.}\label{Fig22}
\end{figure}

\begin{figure}
\center\scalebox{0.4}[0.4]{\includegraphics{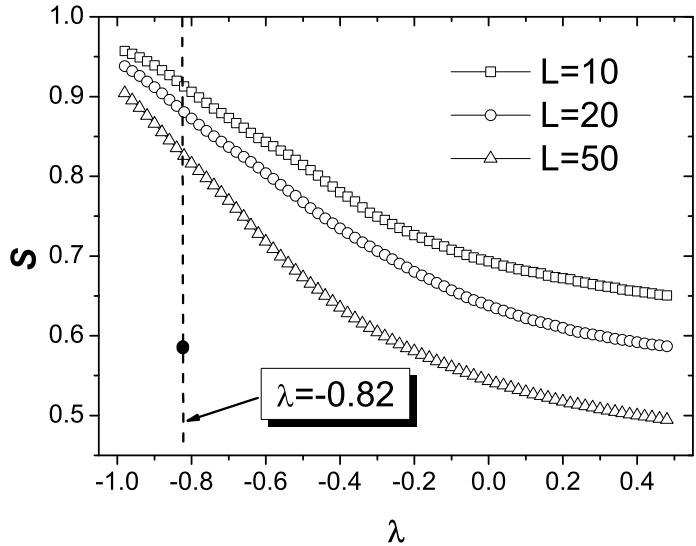}}
\caption{$S$ vs. $\lambda$ when $p=0.9$. Squares, circles and
triangles represent the lengths $L=10, 20$ and $50$, respectively.
The black point ($\bullet$) corresponds to the diversity obtained by
the standard CF with $L=20$. All the data points are averaged over
ten independent runs with different data-set divisions.
}\label{Fig23}
\end{figure}

\section{Improved algorithm}
To investigate the effect of second-order user correlation to
algorithm performance, we use a linear form to investigate the
effect of the second-order user correlation to MCF, where the user
similarity matrix could be demonstrated as
\begin{equation}
\textbf{H} =\textbf{S}+\lambda\textbf{S}^{2},
\end{equation}
where $\textbf{H}$ is the newly defined correlation matrix,
$\textbf{S}=\{s_{ij}\}$ is the first-order correlation defined as
Eq. (2), and $\lambda$ is a tunable parameter. As discussed before,
we expect the algorithmic accuracy can be improved at some negative
$\lambda$.

When $p=0.9$, the algorithmic accuracy curves of \emph{MovieLens}
and \emph{Netflix} have clear minimums around $\lambda=-0.82$ and
$\lambda=-0.84$, which strongly support the above discussion.
Compared with the routine case ($\lambda=0$), the average ranking
scores can be further reduced to 0.0826 (improved 20.45\%) and
0.1436( improved 33.25\%) at the optimal values. It is indeed a
great improvement for recommendation algorithms. Since the data
sparsity can be turned by changing $p$, we investigate the effect of
the sparsity on the two data sets respectively, and find that
although we test the algorithm on two different data sets, the
optimal $\lambda_{\rm opt}$ are strongly correlated with the
sparsity in a uniform way for both \emph{MovieLens} and
\emph{Netflix}. Figure \ref{Fig24} shows that when the sparsity
increases, $\lambda_{\rm opt}$ will decrease, and the improvement of
the average ranking scores will increase. These results can be
treated as a good guideline for selecting optimal $\lambda$ of
different data sets. Figure \ref{Fig22} reports the average degree
of all recommended objects as a function of $\lambda$. One can see
from Fig. \ref{Fig22} that when $p=0.9$ the average object degree is
positively correlated with $\lambda$, thus to depress the influence
of mainstream interests gives more opportunity to the less popular
objects, which could bring more information to the users than the
popular ones. When the list length, $L$, bing equal to 20, at the
optimal point $\lambda_{\rm opt}=-0.82$, the average degree is
reduced by 29.3\% compared with the standard CF. When $p=0.9$, Fig.
\ref{Fig23} exhibits a negative correlation between $S$ and
$\lambda$, indicating that to consider the second-order correlations
makes the recommendation lists more diverse. When $L=20$, the
diversity $S$ is increased from 0.592 (corresponding to the standard
CF) to 0.880 (corresponding to the case $\lambda=-0.82$ in the
improved algorithm). Figure \ref{Fig22} and Figure \ref{Fig23} show
how the parameter $\lambda$ affects the average object degree
$\langle k\rangle$ and diversity $S$, respectively. Clearly, the
smaller $\lambda$ leads to less popularity and higher diversity, and
thus the present algorithm can find its advantage in recommending
novel objects with diverse topics to users, compared with the
standard CF. Generally speaking, the popular objects must have some
attributes fitting the tastes of the masses of the people. The
standard CF may repeatedly count those attributes and assign more
power for the popular objects, which increases the average object
degree and reduces the diversity. The present algorithm with
negative $\lambda$ can to some extent eliminate the redundant
correlations and give higher chances to less popular objects and the
objects with diverse topics different from the mainstream
\cite{NJP}.

\begin{table}
\caption{Algorithmic performance for \emph{MovieLens} data when
$p=0.9$. The precision, diversity and popularity are corresponding
to $L=50$. NBI is an abbreviation of the network-based
recommendation algorithm, proposed in Ref. \cite{Zhou2007}.
Heter-NBI is an abbreviation of NBI with heterogenous initial
resource distribution, proposed in Ref. \cite{Zhou2007b}. CB-CF is
an abbreviation of the correlation-based collaborative filtering
method, proposed in Ref. \cite{Liu2009d}. Improved MCF is an
abbreviation of the algorithm presented in this paper. The
parameters in Heter-NBI and IMCF are set as the ones corresponding
to the lowest ranking scores (for Heter-NBI \cite{Zhou2007b},
$\beta_{\texttt{opt}}=-0.80$; for CB-CF \cite{Liu2009d},
$\lambda_\texttt{opt}=-0.96$; for IMCF,
$\lambda_{\texttt{opt}}=-0.82$). Each number presented in this table
is obtained by averaging over ten runs, each of which has an
independently random division of training set and probe.}
\begin{center}
\begin{tabular} {cccc}
  \hline \hline
   Algorithms     & $\langle r\rangle$  & $S$ & $\langle k\rangle$  \\
   \hline
   GRM & 0.1390 & 0.398 & 259  \\
   CF  & 0.1168 & 0.549 & 246  \\
   NBI & 0.1060 & 0.617 & 233  \\
   Heter-NBI & 0.1010 & 0.682 & 220  \\
   CB-CF & 0.0998 & 0.692 & 218 \\
   IMCF      & 0.0877 & 0.826 & 175\\
   \hline \hline
    \end{tabular}
\end{center}
\end{table}

\section{Conclusions}
In this paper, a modified collaborative filtering algorithm is
presented to improve the algorithmic performance. The numerical
results indicate that the usage of diffusion based correlation could
enhance
the algorithmic accuracy. 
Furthermore, by considering the second-order correlations,
$\textbf{S}^2$, we presented an effective algorithm that has
remarkably higher accuracy. Indeed, when $p=0.9$ the simulation
results show that the algorithmic accuracy can be further improved
by 20.45\% and 33.25\% on \emph{MovieLens} and \emph{Netflix} data.
Interestingly, we found even for different data sets, the optimal
value of $\lambda$ exhibits a uniform tendency versus sparsity.
Therefore, if we know the sparsity of the training set, the
corresponding optimal $\lambda_{\rm opt}$ could be approximately
confirmed. In addition, when the sparsity gets less than 1\%, the
improved algorithm wouldn't be effective any more, while as the
sparsity increases, the improvement of the presented algorithm is
enlarged.

Ignoring the degree-degree correlation in user-object entries, The
algorithmic complexity of MCF is $\textbf{O}(m\langle
k_u\rangle\langle k_o\rangle+mn\langle k_o\rangle)$, where $\langle
k_u\rangle$ and $\langle k_o\rangle$ denote the average degrees of
users and objects. The first term accounts for the calculation of
user correlation, and the second term accounts for the one of the
predictions.  It approximates to $\textbf{O}(mn\langle k_o\rangle)$
for $n\gg \langle k_u\rangle$. Clearly, the computational complexity
of MCF is much less than that of the standard CF especially for the
systems consisted of huge number of objects. In the improved
algorithm, in order to calculate the second-order correlations, the
diffusion process must flow from the user to the objects twice,
therefore, the algorithmic complexity of the improved algorithm is
$\textbf{O}(n\langle k_u\rangle^2\langle k_o\rangle^2+mn\langle
k_o\rangle)$. Since the magnitude order of the object $m$ is always
much larger than the ones of $\langle k_u\rangle$ and $\langle
k_o\rangle$, the improved algorithm is also as comparably fast as
the standard CF.

Beside the algorithmic accuracy, two significant criteria of
algorithmic performance, average degree of recommended objects and
diversity, are taken into account. A good recommendation algorithm
should help the users uncovering the hidden (even dark) information,
corresponding those objects with very low degrees. Therefore, the
average degree is a meaningful measure for a recommendation
algorithm. In addition, since a personalized recommendation system
should provide different recommendations lists according to the
user's tastes and habits \cite{Ecommerce2001}, diversity plays a
crucial role to quantify the personalization \cite{Ziegler,PNAS}.
The numerical results show that the present algorithm outperforms
the standard CF in all three criteria. How to automatically find out
relevant information for diverse users is a long-standing challenge
in the modern information science, we believe the current work can
enlighten readers in this promising direction.

How to automatically find out relevant information for diverse users
is a long-standing challenge in the modern information science, the
presented algorithm also could be used to find the relevant
reviewers for the scientific papers or funding applications
\cite{Liu1,Liu2}, and the link prediction in social and biological
networks \cite{Zhou2009,Lu2009}. We believe the current work can
enlighten readers in this promising direction.

We acknowledge GroupLens Research Group for providing us the data.
This work is partially supported by and National Basic Research
Program of China (No. 2006CB705500), the National Natural Science
Foundation of China (Nos. 10905052, 70901010, 60744003), the Swiss
National Science Foundation (Project 205120-113842), and Shanghai
Leading Discipline Project (No. S30501). T.Z. acknowledges the
National Natural Science Foundation of China under Grant Nos.
10635040 and 60973069.

\end{document}